# 40 Gbps Readout interface STARE for the AGATA Project

N. Karkour[1], V. Alaphilippe[1], J. Collado[7], N. Dosme[1], L. Gibelin[1], V. Gonzalez[2], X. Grave[1], J. Jacob[1], X. Lafay[1], E. Legay[1], D. Linget[1], A. Pullia[4], M. quenez[3], D. sidler[5], N. Tessier[3], G. Vinther-Jørgensen[6],

(1) Université Paris-Saclay, CNRS/IN2P3, IJCLab, 91405 Orsay, France

(2) Departamento de Ingeniería Electrónica, Universitad de Valencia, Valencia, Spain

(3) ESME (Ecole Supérieure Mécanique et Electronique) Sudria, Paris, France

(4) INFN, Milan, Italy

(5) Dept. of Computer Science, ETH Zürich

(6) Technical University of Denmark, Lyngby, Denmark

(7) Instituto de Física Corpuscular, CSIC, Paterna, Valencia, Spain

*Abstract–* The Advanced GAmma Tracking Array (AGATA) multi detector spectrometer will provide precise information for the study of the properties of the exotic nuclear matter (very unbalanced proton (Z) and neutron (N) numbers) along proton- and neutron- drip lines and of super-heavy nuclei. This is done using the latest technology of particle accelerators. The AGATA spectrometer consists of 180 high purity Germanium detectors. Each detector is segmented into 38 segments. The very harsh project requirements are to measure gamma ray energies with very high resolution (< 1x 10 -3) at a high detector counting rate (50 Kevents / sec / crystal). This results in a very high data transfer rate per crystal (5 to 8 Gbps). The 38 segments are sampled @ 100 MHz with 14 bits of resolution. The samples are continuously transferred to the CAP (Control And Processing) module which reduces the data rate from 64 Gbps to 5 Gbps. The CAP module also adds continuous monitoring data which results in total outgoing data rate of 10 Gbps. The STARE module is designed to fit between the CAP module and the computer farm. It will package the data from the CAP module and transmit it to the server farm using a 10 Gbps UDP connection with a delivery insurance mechanism implemented to ensure that all data is transferred.

## I. Introduction

AGATA is designed to provide precise information for the study of the properties of the exotic nuclear matter along proton- and neutron- drip lines and of super-heavy nuclei. The AGATA spectrometer consists of 180 high purity Germanium detectors. The AGATA project [1] instrumentation has been updated through 3 phases (Phase0, Phase1 and Phase2). This actual instrumentation is part of the phase 2 electronics. Fig. 1 shows the different phases of AGATA from Legnaro (phase 0) to GANIL (phase 1). Each detector is segmented into 38 segments. Each segment is preamplified and the output is sampled @ 100 MHz with 14 bits of resolution using Flash Analog to Digital Converters (FADC) boards called the DIGIOPT12 [8]. The total raw data rate per detector is around 64 Gbps (1,6 Gbps x 38 segments plus additional diagnostic and identification information of 2 x 1,6 Gbps). The raw data is processed in real time, compressed to 10 Gbps and sent to the STARE (Serial Transfer Acquisition and Readout over Ethernet) module. The general overview of the AGATA detector acquisition is shown in Fig. 2. In this article we will report on the STARE prototype hardware and firmware design. We will describe the R&D phase for the Proof of Concept. This includes Hardware design, Firmware Design, the results and a description of the future work.

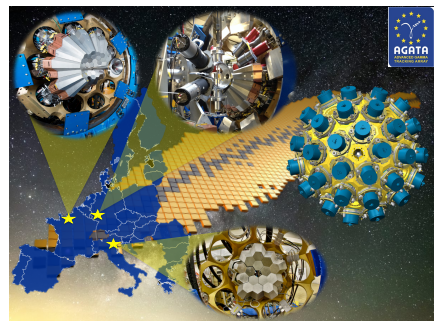

Fig. 1. AGATA Phase 0, 1 and 2 detector design.

The STARE module design is an FPGA Mezzanine Card (FMC) standard module which will be mounted on a PACE module (Preprocessing And Control through Ethernet). The PACE [4] module is a collaboration between the IFIC

(Valencia Spain), the Department of Electronic Engineering (Valencia, Spain), CNRS/IN2P3/IJCLAB and INFN Milan.

The construction of STARE is fully financed by the IN2P3 (Institut Nationale de Physique Nucléaire et de Physique de Particule) institute of the CNRS (Centre National de Recherche Scientifique) in France through the AGATA MOU (Memorandom Of Understanding)

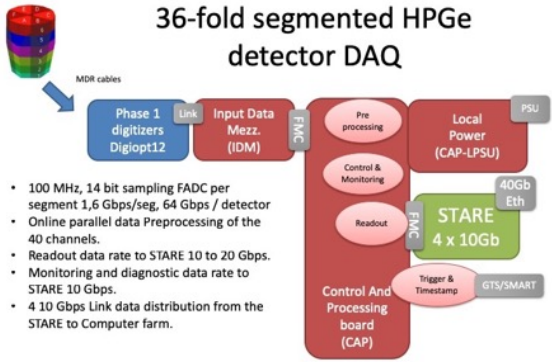

Fig. 2. Instrumentation of an AGATA detector. PACE block diagram and specifications.

## II. R&D PRELIMINARY PHASE

The R&D phase started in 2017, looking for a 10 Gbps ethernet IP design compatible with Xilinx FPGAs. After gathering and sourcing different solutions (costs up to 150 k€!) a collaboration was established with the ZTH (Zurich University) to integrate a 10 Gbps IP developed at the university with the VC709 Xilinx evaluation board. This IP[2] is an open source IP available to all research institutions and is developed by David Sidler [6] who kindly helped with the integration. The first validation was based on a simple data generator IP that could transfer data at various transfer rates and on the server side the IPERF [7] program was used to receive the data from the VC709 10 Gbps UDP port. This program helped us to validate the IP design with the AGATA data transfer constraints. The first results in Fig. 3 show 8,6 Gbps transfer rate from the VC709 to the computer server using UDP (User Datagram Protocol). This test convinced us to proceed with the R&D phase. The results below showed that at 5 Gbps the data loss is very low and hence it is very promising. At a rate of 8.6 Gbps the data loss was important, and so it was decided to have multiple UDP transceivers so that the data loss is 0% as it is in the AGATA design requirements. Moreover, the Virtex 7 FPGA is a small FPGA to house the 4 UDP transceivers and the high speed clock connection used for the transceivers is not flexible. Because of this, it was decided to design the Proof of Concept using the KU040 FPGA integrated in the KCU105 development board.

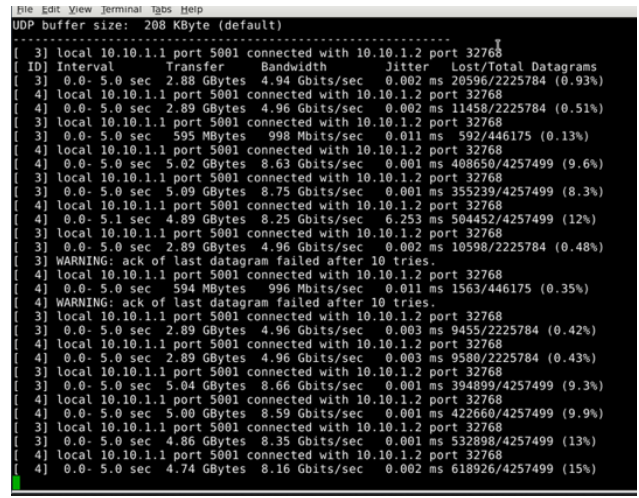

Fig. 3. R&D results using ZTH UDP IP, VC709 and the IPERF program.

## III. THE PROOF OF CONCEPT

The POC was made to validate the complete instrumentation chain of 1 AGATA channel using 1 DigiOpt12 (FADC module), the IFIC-UV PACE Preprocessing FPGA R&D evaluation board (to record traces from the DIGIOPT12 and send them to the STARE), and the IJCLAB STARE R&D evaluation board the KCU105. The Server STARE01 is used to receive the STARE data. Fig. 4 shows the POC set-up block diagram. Fig. 5 shows the POC modules all together installed in IJCLAB.

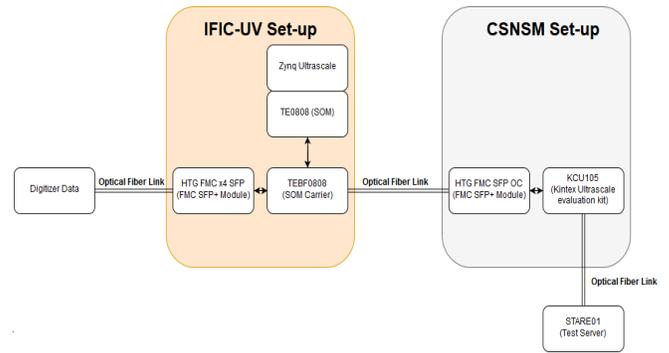

Fig. 4. Block Diagram of the POC of AGATA instrumentation electronics.

The STARE firmware design was based on the following characteristics :

1. UDP IP provided by David Sidler (ETH Zurich)[2] and IPBus firmware from CERN[2]
2. Event splicing/reconstruction
3. 1 UDP transceiver out of 4 programmed
4. Package loss simulation

Fig. 5. Photos of the POC of AGATA electronics. Top CAP and DIGIOPT12, bottom STARE modules

There were several tests made to validate the AGATA POC. The first test was to check the STARE UDP transmission using a data generator implemented in the STARE Firmware. The second test was to connect the IFIC evaluation board to the STARE without the DIGIOPT module and transfer data from a data generator implemented in the IFIC board firmware. The third test was to connect all modules. The tests were made successfully between the 3 modules. The server software used to validate the data quality was only made to work for the first and second tests. Moreover 2 types of receiving data programs were used on each test. The first program was receiving data but not writing to disk. Fig. 6 shows the transfer rate @ 5Gbps with 0% data loss. The Sender was transmitting 2KB of event data @300 Kevents/sec. The second program was receiving data and writing to disk. We can see how in Fig. 7 the transfer rate is degraded as the program was not able to store the data as quickly as it arrived. This led to data loss and the transfer rate quickly degraded. It was important to identify the constraints source that comes from the computer program optimization and the server calculating power. It confirmed that either a more optimized software, the use of multiple 10 Gbps UDP ports to multiple servers or 1 UDP port to multiple servers is required. Another idea that arose from this test was to develop an IP guaranteeing reliable data transfer at 10Gbps. This will be done in future firmware upgrades.

Fig. 6. STARE Transfer rate without writing to disk. No data loss

Fig. 7. STARE Transfer rate while writing to disk. Data loss starts occurring after a few transferred buffers.

IV. STARE HARDWARE DESIGN

The STARE is part of the Preprocessing electronics of the phase2 electronics for the AGATA project. It is an FMC board that will be mounted on the PACE module. The AGATA Germanium detector signals are digitized in the DIGIOPT modules and the continuous data is processed inside the PACE module. The Readout manager of the PACE module is based on Xilinx AXI stream AURORA interface. It sends data to the STARE and the STARE processes the received data at up to 10 Gbps in order to transfer it to the server farm using the UDP protocol. The STARE module is equipped with external memory to allow retransmission of any packages not correctly received by the server farm. Fig. 8 shows the STARE conceptual design and philosophy of work. It can handle up to 4 independent 10 Gbps transceiver lines on its input and send data to 4 or more independent 10 Gbps servers or computer farm. One UDP channel can send data to different servers, and the STARE has 4 independent UDP channels

Fig. 8. STARE Conceptual design

The STARE board is composed of a SOM (System On Module), an FMC connector and four SFP+ 10 Gigabits transceivers. It contains the following components: 2 EEPROMs, a JTAG interface, an external power input, a high precision oscillator and clock generator.

The STARE design respects all electrical requirements described in the FMC Vita standard [9]. It does not fully comply with the mechanical requirements in the Vita FMC standard because of the SOM and the 4 SFP+ connectors size, which makes the height of the STARE over 10 mm which is the requirement stated in the Vita FMC standard. The length of the STARE board will also be longer than the VITA recommended rules. The PCB design will have a rugged metallic border to insure thermal conductivity since the SOM and SFP transceivers temperature might increase during full speed transmission.

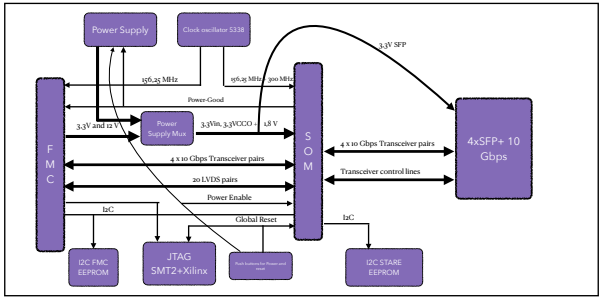

Fig. 9. STARE electrical and interconnection block diagram.

### A. Prototype board design

The STARE board prototype has some functions which will not be used in the production board. The power supply management on the STARE board is very complex and mandatory because the power up sequence of the SOM module and the SFP+ transceivers is a challenge. This is because the carrier boards used to test the STARE board might not deliver the necessary power supply at the first power up stage of the STARE prototype. Fig. 10 shows the block diagram of the STARE power supply management and power up sequence. Moreover, the JTAG programming will use different modes of connections (FMC JTAG, simple Xilinx HX Programmer or a JTAG SMT2 module). In the production phase only the FMC JTAG programming will be implemented. This extra functionality will make the STARE prototype board longer than the production board but will make development and debugging significantly easier. Fig. 11 shows the STARE prototype layout.

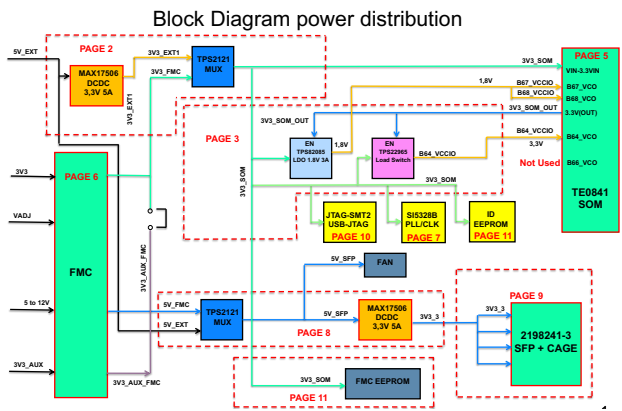

Fig. 10. STARE power supply management and secure Power up sequence

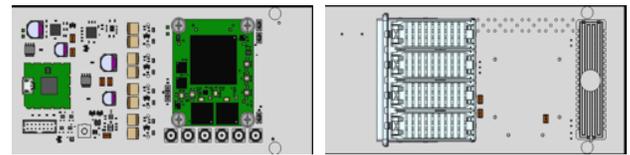

Fig. 11. STARE Prototype Layout :Top side (left) and bottom side (right)

### B. Production board design

The STARE production board will be smaller than the prototype because all onboard JTAG connections and the Power supply Multiplexers will be removed. Fig. 12 shows the production version of the STARE board layout.

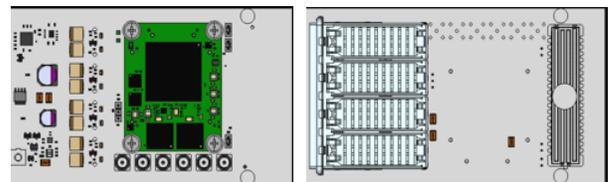

Fig. 12. STARE Production Layout :Top side (left) and bottom side (right)

### C. SOM module

The SOM which will be used on the STARE module is the TE0841 TRM (SOM) from Trenz. It requires 50 × 40 mm of space and 50 mm height. The SOM is connected through two LSHM-150-04.0-L-DV-A-S-K-TR (100 pins) connectors and one LSHM-130-04.0-L-DV-A-S-K-TR (60 pins) connector to the STARE. There is 2GB of external DDR memory used to ensure that lost data can be retransmitted.

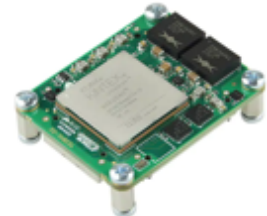

Fig. 13. Trenz SOM module TE0841 (photo from Trenz website)

### D. FMC interface

The STARE is mounted on the PACE board through an FMC HPC (High Pin Count) connector. This connector will contain four 10 Gbps transceiver data pins which will be connected to one of the FPGAs transceiver banks. The GBT-M2C clk0 clock pins will be connected to the 156.25 MHz clock output of the 5338 programmable PLL to propagate the clock signal from the STARE board to the carrier card.

The FMC standard requires a single connector board to have a fixed width of 69mm. The HPC connector needs a space of 14.48mm x 53.90mm. 2 holes will be mounted on each side of the FMC connector to fix the STARE on the PACE board through spacers so that the FMC connector does not support the mechanical stress of the STARE. There are 20 LVDS pair signals connected between the STARE SOM and the PACE FPGA for future use (like the IPBus interface or any other control lines). The clk0-M2C-CC signal will connect a clock signal from the SOM to the PACE FPGA. All power and power control pins respect the VITA standard. A set of Reset signals is connected to manage Hardware and Software Resets on the STARE. These will be used for initialization procedures. Because the STARE board contains several high-speed computing technologies and programmable components the power supply sequence and the initialization procedure are very complex. The high-speed links and data management is extremely complex in the case of the AGATA instrumentation system and the data acquisition software. Any desynchronization in the system clock, memory buffers, system counters or online preprocessing steps will lead to data recording errors and system hangout.

## V. STARE FIRMWARE DESIGN

The STARE firmware is targeted at the KU040 FPGA from Xilinx Corporation. The R&D phase started using the VC709 Virtex 7 technology from Xilinx. The first firmware was developed inside the VC709 to validate the conceptual design of the STARE, using 10 Gbps transfer results over UDP FPGA Network Stack developed by David Sidler from ETH Zurich. The Virtex 7 showed immediately its technology limitations to fulfil the AGATA project requirements. The POC firmware was developed using the KCU105 development board sold by Xilinx. This development board is based upon a modern Kintex Ultrascale (KU040) FPGA and hosts a number of connectors used for high speed data transfer to and from the FPGA.

The POC firmware was very simple it consists of a dummy data generator, an IPBus interface firmware from CERN [2] and the FPGA Network Stack developed by David Sidler from ETH Zurich [3]. These three objects were combined into a single project. The IPBus is providing slow control and debugging functionality via a 1Gbps Ethernet connection while the FPGA Network Stack provides a 10Gbps Ethernet link supporting UDP [6].

The merging of these two code bases resulted in a fairly complex project. As the FPGA Network Stack and IPBus had to share FPGA clocking primitives for their high-speed transceivers a good code structure wasn't initially achieved. This resulted in huge VHDL files and a project that was hard to make sense of.

To make the project more accessible the structure shown in Fig. 14 was developed. Based on the experiences from the initial implementation a number of improvements could provide a much cleaner code structure making the project easier to manage. While restructuring the code the IPBus slow control functionality should also to be implemented. As shown on Fig. 14. The STARE firmware is divided into 4 modules : AURORA interface, The Data manager, the IPBus interface and the UDP interface [5]. Each module contains an IPBus slave providing slow control and status monitoring. The Aurora interface receives data from the PACE module FPGA through Gigabit Transceivers. These transceivers use a 156.25 MHz clock to achieve 10 Gbps. The data manager takes the event data received and splice it into small packets to be sent over UDP. The UDP interface adds specific protocol data to each packet and transmits the packet to the server.

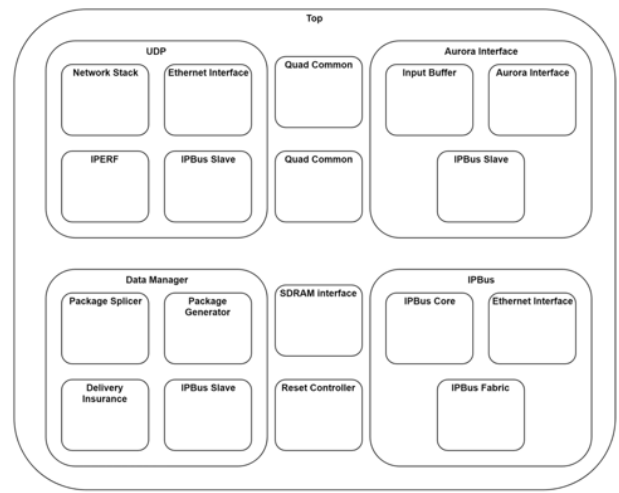

Fig. 14. STARE firmware block diagram

### A. Reliable UDP 10 Gbps transfer SRP (Selective Repeat Protocol)

One of the requirements for the STARE board is that all the data received from the pre-processing must be delivered to the server. As the transport protocol used is UDP, this functionality is not guaranteed by default and must therefore be added. The other widely used transport protocol TCP does provide guaranteed delivery. Thus, an explanation for why UDP was chosen anyway is needed. As described in [3] the FPGA Network Stack provided by David Siddler did originally contain an implementation of both UDP and TCP. Both interfaces were tested extensively by another visitor in the early phase of the STARE R&D phase. Here it was discovered that the TCP implementation could not provide more than about 2Gbps of bandwidth on a single TCP session. The reason for this can be found in the documentation for the FPGA Network Stack [3]. Here the stated goal is to develop a TCP implementation with the capability of handling several TCP sessions at the same time efficiently. This effectively means that the TCP implementation is optimised for a completely different use case than that of the STARE board. The STARE board will only have one session between the board and a server and would therefore never utilise the full 10Gbps bandwidth. This leaves the UDP interface as the only feasible option. The UDP interface is able to utilise the full bandwidth for a single session. However, the downside to this is that a protocol is needed on top of the UDP protocol to guarantee delivery. This will slightly decrease the available bandwidth and require time to be implemented. Implementing a reliability mechanism for UDP is something that has been done before. In cases where TCP is too resource heavy to implement, a UDP connection with a simple reliability protocol on top can often be the solution. When deciding on a protocol, it is important to define what exactly is expected. For the STARE project, the requirement is that all data is delivered and delivered exactly once. This means that lost data should be re-sent and data that is sent twice should be discarded. Several protocols were studied and the chosen protocol was the Selective Repeat Protocol which is the more efficient option. Fig. 15 shows the block diagram of the Selective Repeat protocol.

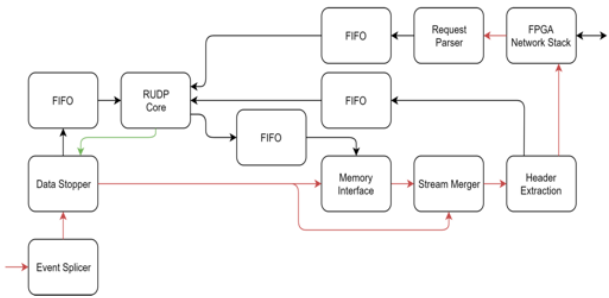

Fig. 15. Selective Repeat Protocol block diagram.

This however comes at the cost of increased complexity. The operation of the protocol is shown on Fig. 16. When using the Selective Repeat protocol, only the lost frame will have to be sent back. This results in a better utilisation of the available bandwidth, especially if the loss rate is high.

It also means that the memory interface must allow single frames to be read back for re-transmission. As each frame is treated individually, it also means that a timeout mechanism for each frame is needed. This greatly increases the implementation complexity on the sending side.

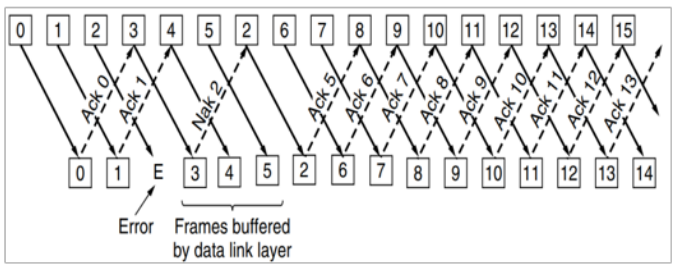

Fig. 16. Selective Repeat Protocol mechanism with intelligent Handshake process, to minimize time shift, and synchronizes quickly.

The core of the Selective Repeat protocol implementation is the RUDP Core. RUDP is short for Reliable UDP which is what the Selective Repeat protocol provides. The RUDP Core keeps track of all the active frames and generates timeouts used to re-send frames in case an acknowledge is not received.

When a frame is allowed to pass through the data stopper the frame is marked as active in the RUDP Core. The next step in the frame lifecycle is for the timeout countdown to start. This happens when the frame has passed through the Header Extraction module. In this way the timeout won't be affected by frames read back from the memory delaying the frame transmission.

The timer is now running. If the acknowledgement frame is received before the timeout occurs the frame is marked as inactive and the RUDP Core is finished with the frame. If the acknowledgement frame isn't received in time the frame times out. When this happens, a request is sent to the Memory Interface block triggering the frame to be read from memory. When the frame passes through the Header Extraction module the timeout is started again. This process will continue until the acknowledgement frame is received. The flow described is shown in Fig. 17.

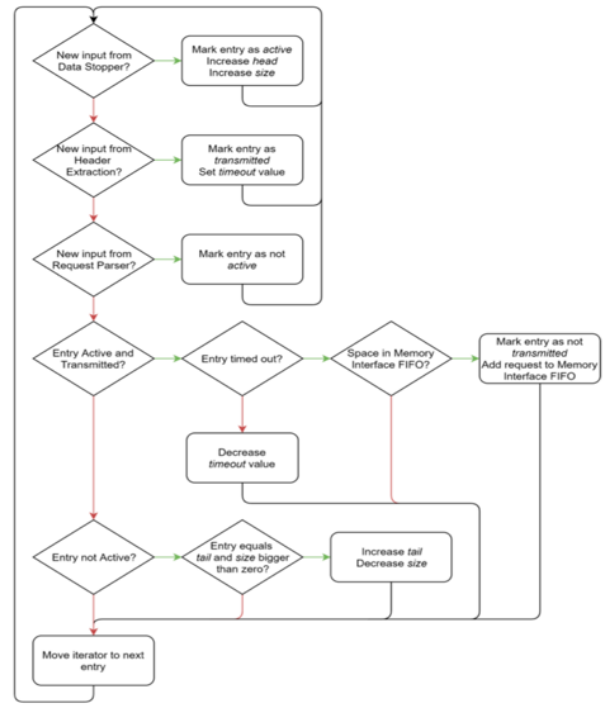

Fig. 17. RUDP core mechanism.

### A. Memory interface

The Memory interface is a very complex block designed in the STARE firmware. The reason for the complexity is that it must be possible to read back frames for retransmissions while still allowing the incoming data to be stored at up to 10 Gbps. The POC Firmware can only store data from a single 10 Gbps input. With the Trenz SOM module it might be possible to handle two 10 Gbps inputs depending on the network latency and memory bandwidth. The memory management block diagram is shown in Fig. 18.

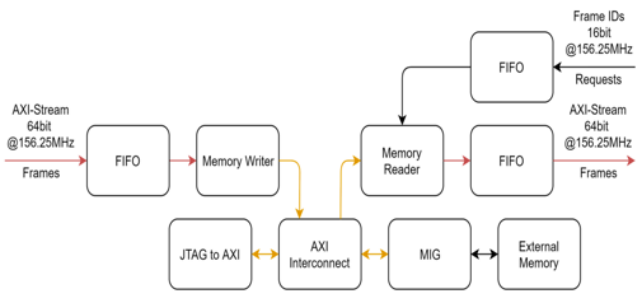

Fig. 18. Memory interface block diagram.

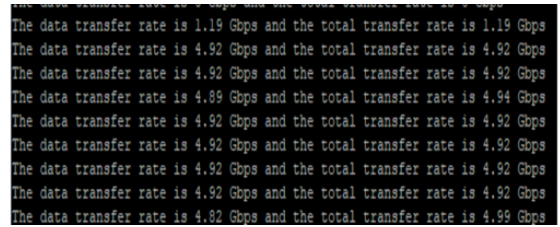

Fig. 19. The output from the software for the Selective Repeat Protocol.

## VI. RESULTS

The implementation of the Selective Repeat Protocol proved more complex than expected. The amount of code and complexity required for solving this fairly simple task was quite high. The result is however a nicely structured and efficient implementation. There are a lot of possible improvements and optimisations for the future. A more rigorous testing and qualification is also needed.

The implementation around the RUDP Core has also proved to be smart. As the RUDP Core has all the information about each frame and the status of the window overall, a lot of functionality can be added. An example could be a mechanism for stopping data transfer if the loss rate gets too high.

The adaptation of Vivado HLS for the implementation has been a big success. Compared to the experience of developing the original Event Splicer the development process using Vivado HLS is much more efficient. Instead of focusing on making an AXI Stream interface function the focus can be squarely on the functionality of the module. The possibility of testing the code with C++ testbenches also helped discover and remove bugs before the module was implemented on the FPGA.

The implemented server software is in no way optimized but will prove useful when testing the firmware. An optimization of the way memory is handled would however be a good idea as a lot of resources are wasted on that. Furthermore, the implementation of checks on the whole frame and not just the header is necessary if the software is to be used for the final qualification of the firmware.

The output from the software can be seen on Fig. 19. This output shows a test where the KCU105 was sending data at 5Gbps with no simulated loss. The event size was set to 8192B and the frame size was at 1400B. This clearly shows that the minimum requirement for the protocol is met.

Attempts were also made to increase the bandwidth further. Unfortunately the implemented Data Generator has no steps between 5Gbps and 10Gbps. When tested at 10Gbps the software only reported a total transfer rate of around 6Gbps. At the same time the monitors in the firmware reported that the full link capacity was used. The reason for this is most likely that the server is unable to process the incoming data.

## VII. FUTUR WORK AND OUTLOOK

The STARE Hardware prototype is under production, 10 modules will be delivered Q4 2020 with qualification milestone with the PACE module in Q1 2021 and with the AGATA detector Q2 2021. Production will be delivered in Q4 2021. The firmware will be implemented inside the SOM modules using the same code as the KCU105 FPGA since both FPGAs are the same and the external DDR memory interface is the same. The firmware will be qualified in Q1 2021 and the upgrades will continue until Q2 2022.

## ACKNOWLEDGMENT

We would like to thank all the people involved in this project. Thanks to our financial departments institutes who supported our R&D development (IN2P3). I would like to thank first David Sidler from the ZTH Zurich (now he is working at Microsoft company) without him we wouldn't validate the preliminary R&D phase. I would like to thank Andrea Triossi to the great work he has done to help us progressing in our R&D phase. I would like to thank my 2 excellent students from ESME Sudria university Melissa Quenez and Nicolas Tessier. I would like to give a great thanks to Gustav my student and friend who put the necessary ingredients to make the Proof of concept possible and who designed the Selective Repeat Protocol. His contribution is extremely valuable. My great thanks to all my technical staff who has invested a lot of time and effort in this project. Thanks also to the scientific staff, who helped us to achieve our goals in this project. Many thanks to all the AGATA collaborators especially our Valencia team and Milano team who design the PACE module.

## REFERENCES

[1] S. Akkoyun, A. Algora, B. Alikhani, F. Ameil et al., "Agata—advanced gamma tracking array," Nuclear Instruments and Methods in Physics

Research Section A: Accelerators, Spectrometers, Detectors and Associated Equipment, vol. 668, pp. 26 – 58, 2012.

[2] C. Ghabrous Larrea, K. Harder, D. Newbold, D. Sankey, A. Rose, A. Thea, and T. Williams, "IPbus: a flexible Ethernet-based control system for xTCA hardware," JINST, vol. 10, no. 02, p. C02019, 2015

[3] D. Sidler, G. Alonso, M. Blott, K. Karras et al., "Scalable 10Gbps TCP/IP Stack Architecture for Reconfigurable Hardware."

[4] J. Collado *et al.*, "A new preprocessing and control board for the phase 2 electronics of AGATA experiment," *2016 IEEE-NPSS Real Time Conference (RT)*, Padua, 2016, pp. 1-4, doi: 10.1109/RTC.2016.7543161.

[5] G. Vinther-Jørgensen, "Internship report 2," 2019

[6] D. Sidler, "In-network data processing using fpgas," Ph.D. dissertation, 2019-09

[7] Cottrell, Les. "Measuring End-To-End Bandwidth with Iperf Using Web100." (2003). DOI:10.2172/813039

[8] A. Pullia, D. Barrientos and S. Capra, "Open-source diagnostic tool with GUI for the new AGATA/GALILEO/EUCLIDES digitizer cards," 2014 IEEE Nuclear Science Symposium and Medical Imaging Conference (NSS/MIC), Seattle, WA, 2014, pp. 1-5, doi: 10.1109/NSSMIC.2014.7431138.

[9] ANSI/VITA 57.1 FPGA Mezzanine Card (FMC) Standard.